\documentclass[times]{qjrms3}

\usepackage{natbib}

\def\volumeyear{2007}
\def\volumenumber{133}
\def\DOI{qj.179}

\begin{document}

\runningheads{I. A. Boutle et al.}{Boundary-layer friction in cyclones}

\title{A note on boundary-layer friction in baroclinic cyclones}

\author{I. A. Boutle\affil{a}\corrauth,
R. J. Beare\affil{b}, S. E. Belcher\affil{a} and R. S. Plant\affil{a}}

\address{\affilnum{a}Department of Meteorology, University of Reading, UK\\
\affilnum{b}Met Office, Exeter, UK}

\corraddr{I. A. Boutle, Department of Meteorology, University of Reading, PO Box 243, Reading, Berkshire, RG6 6BB, UK.\\E-mail: i.a.boutle@reading.ac.uk}

\begin{abstract}
The interaction between extratropical cyclones and the underlying boundary layer has been a topic of recent discussion in papers by \cite{AdaBHP06} and \cite{Bea07}. Their results emphasise different mechanisms through which the boundary layer dynamics may modify the growth of a baroclinic cyclone. By using different sea-surface temperature distributions and comparing the low-level winds, the differences are exposed and both of the proposed mechanisms appear to be acting within a single simulation.
\end{abstract}


\received{22 May 2007}
\revised{1 October 2007}
\accepted{3 October 2007}

\maketitle

\section{Introduction}

Recent papers by \cite{AdaBHP06} and \cite{Bea07} have provided two
differing explanations for the way in which the boundary
layer affects dry, baroclinic cyclones. \cite{AdaBHP06} provided a new
mechanism for the frictional spin-down of an extratropical cyclone,
based on baroclinic potential vorticity (PV) generation. PV generated in the boundary layer (due to a component of stress anti-parallel to the tropospheric thermal wind) is advected along the warm conveyor belt and vented from the boundary layer. It turns cyclonically and accumulates above the low centre as a positive PV anomaly of large horizontal extent, but trapped between isentropes in the vertical. This shape is associated mainly with increased static stability, inhibiting communication between the upper- and lower-level features of the developing wave, and so reducing growth.

However, \cite{Bea07} showed that the dominant low-level PV anomaly in his model was not associated with the spin-down of the cyclone, and that the region of greatest surface stress was most important in restricting cyclone growth. This region, to the west of the surface low centre, is characterised by a well-mixed boundary layer, and so is not associated with a significant contribution to the PV budget. Instead, its effects are thought to occur via the Ekman pumping mechanism -- the additional stress enhances convergence of near surface winds, which must create upward vertical motion due to continuity. This reduces cyclone growth by vortex squashing in the interior. 

This note aims to clarify the differences between the two studies and demonstrate that both mechanisms can be seen in the same simulation.

\section{Methodology}

To mimic the large-scale atmospheric structure, both previous studies specified a mid-latitude jet with a potential temperature profile calculated from thermal wind balance. However, \cite{AdaBHP06} took their sea-surface temperature field to be identical to that of the first model level at the initial time, whilst \cite{Bea07} chose a constant sea-surface temperature across the entire domain.

The cyclone initialisation also differed between the studies. \cite{AdaBHP06} followed the same initialisation as for the lifecycle denoted LC1 by \cite{ThoHM93}, in which a normal mode perturbation is added to the basic state, evolving slowly into a cyclone over $5$--$10$ days. The study of \cite{Bea07} included a finite-amplitude upper-level vortex in the initial conditions, triggering rapid cyclogenesis over $2$--$3$ days, and showing much similarity to the conceptual model of \cite{ShaK90}. Therefore, the experiments represent different realisations of the spectrum of real-world cyclones, and neither can claim to be closer to reality than the other. Here we consider an intermediate experiment, repeating the simulations of \cite{Bea07}, but choosing our sea-surface temperature in the same manner as \cite{AdaBHP06}.

\section{Results}

\begin{figure*}[tbhp]
\centering
\includegraphics[width=\columnwidth]{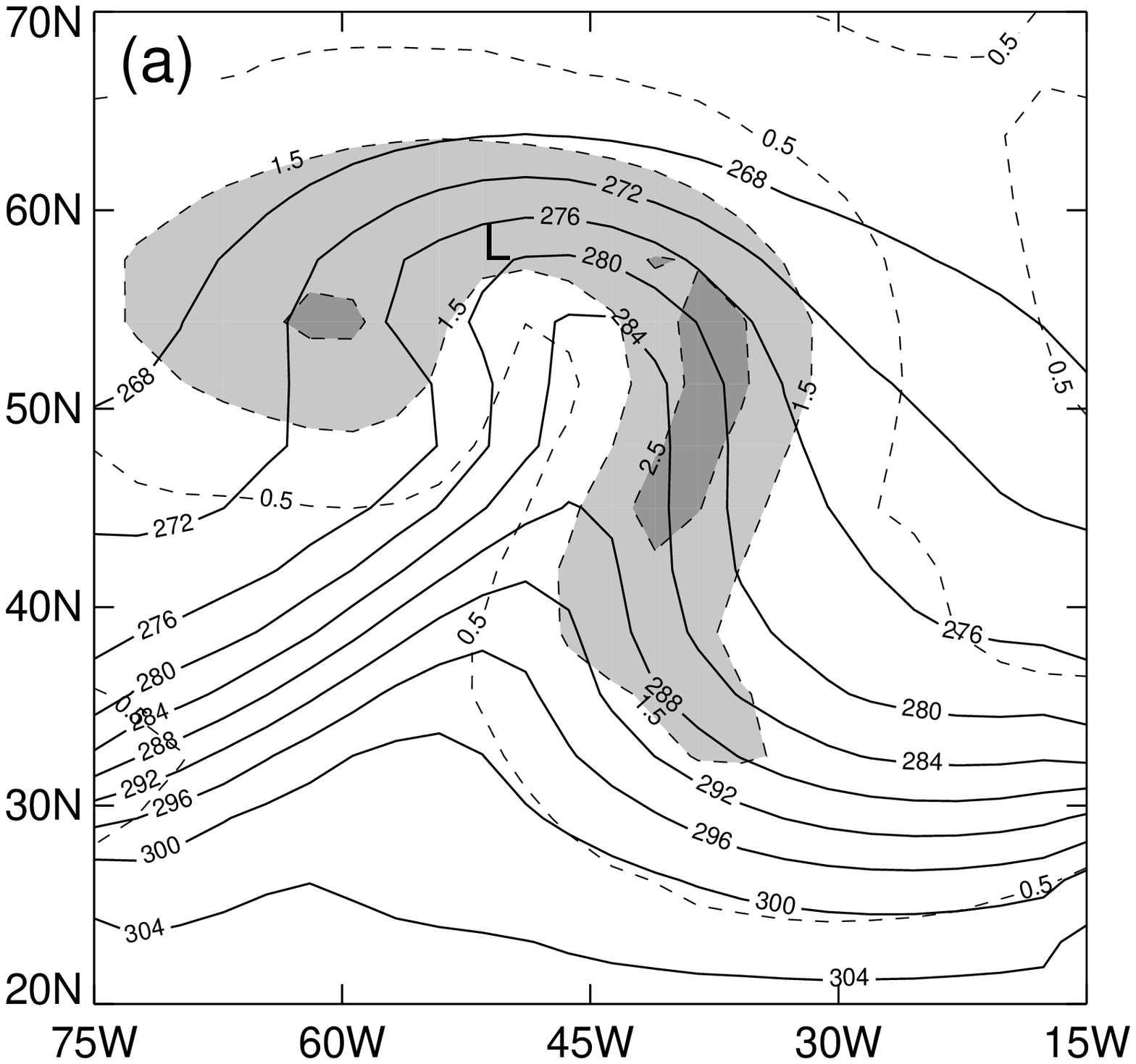}
\includegraphics[width=\columnwidth]{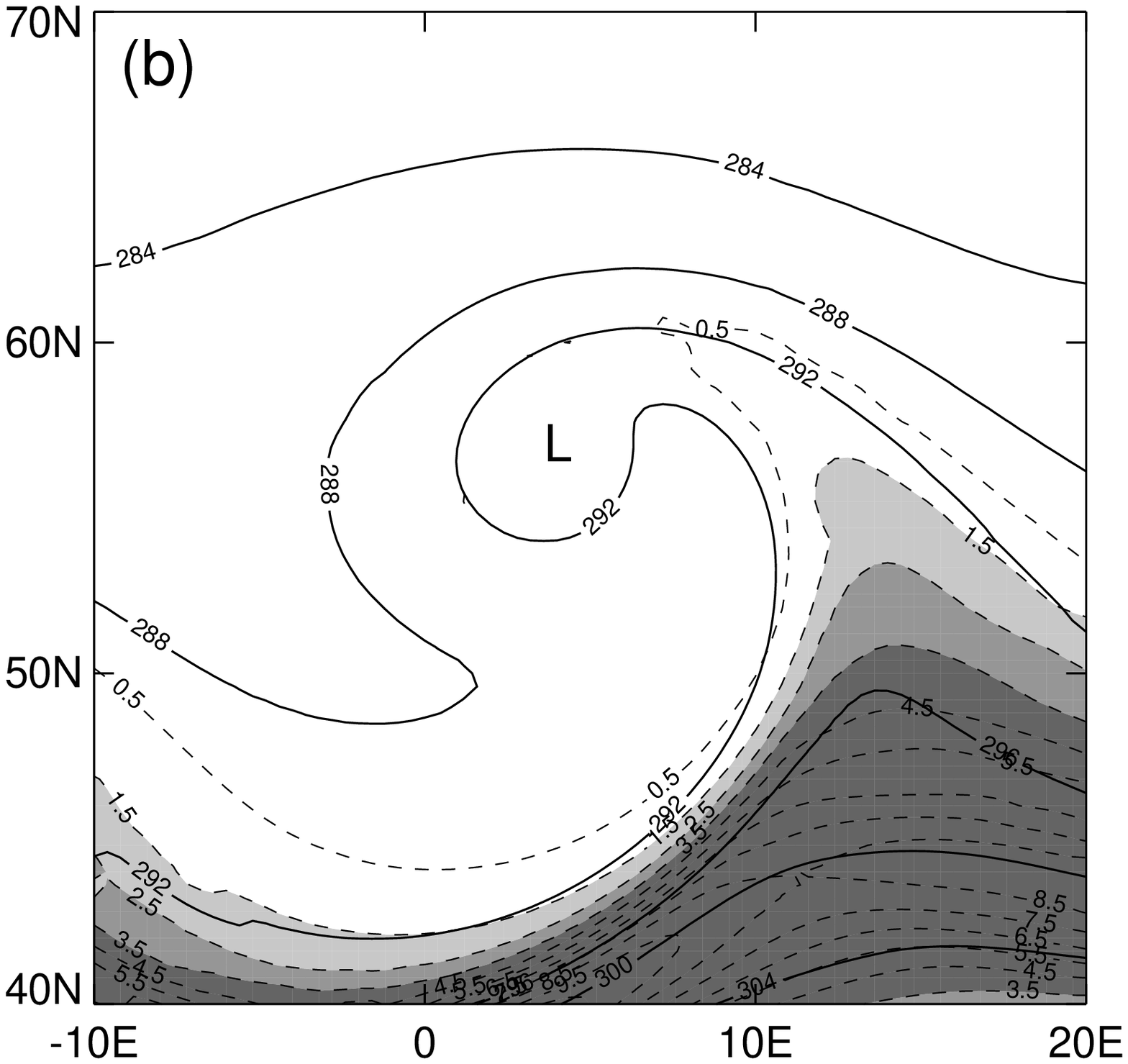}
\includegraphics[width=\columnwidth]{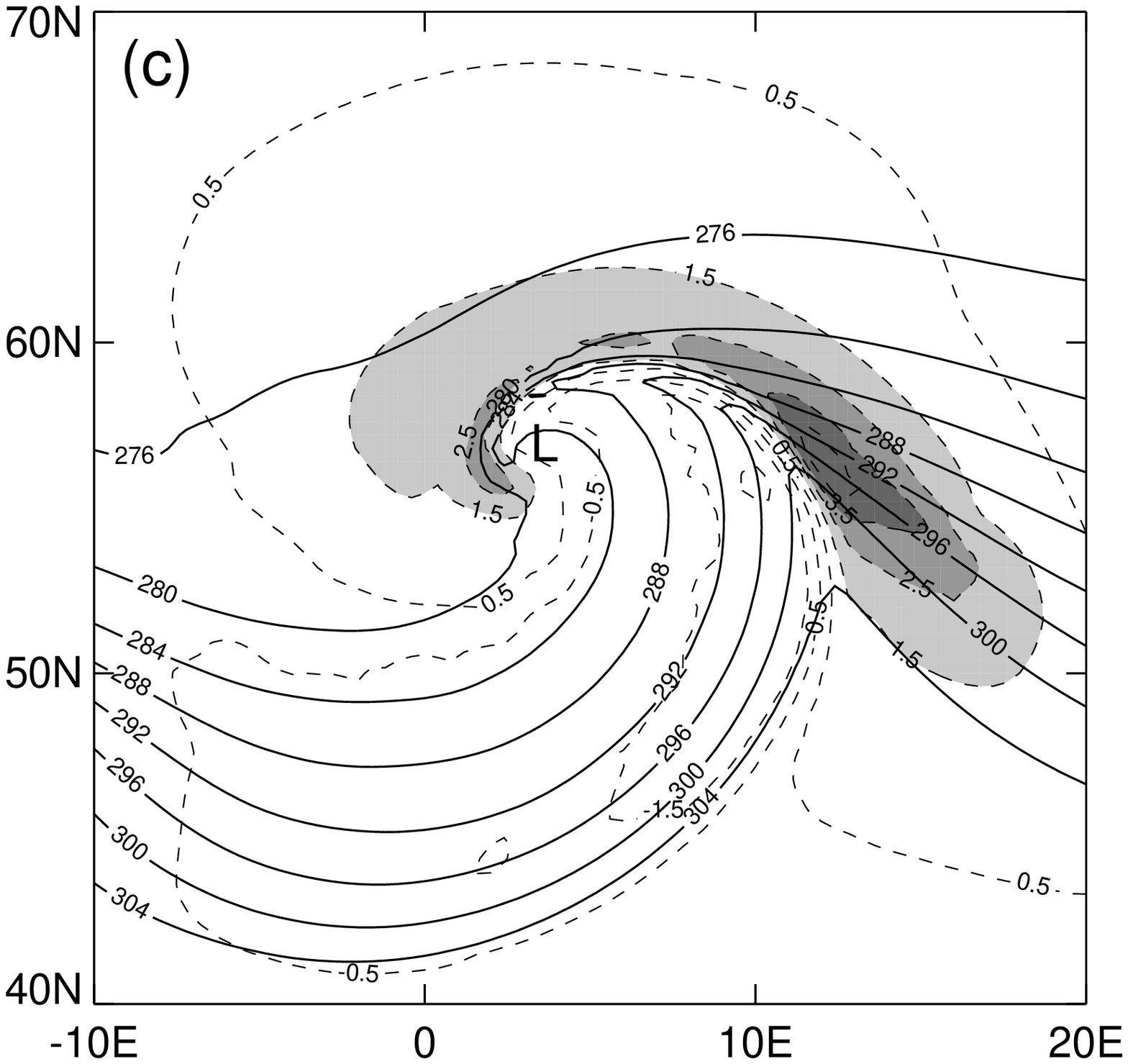}
\includegraphics[width=\columnwidth]{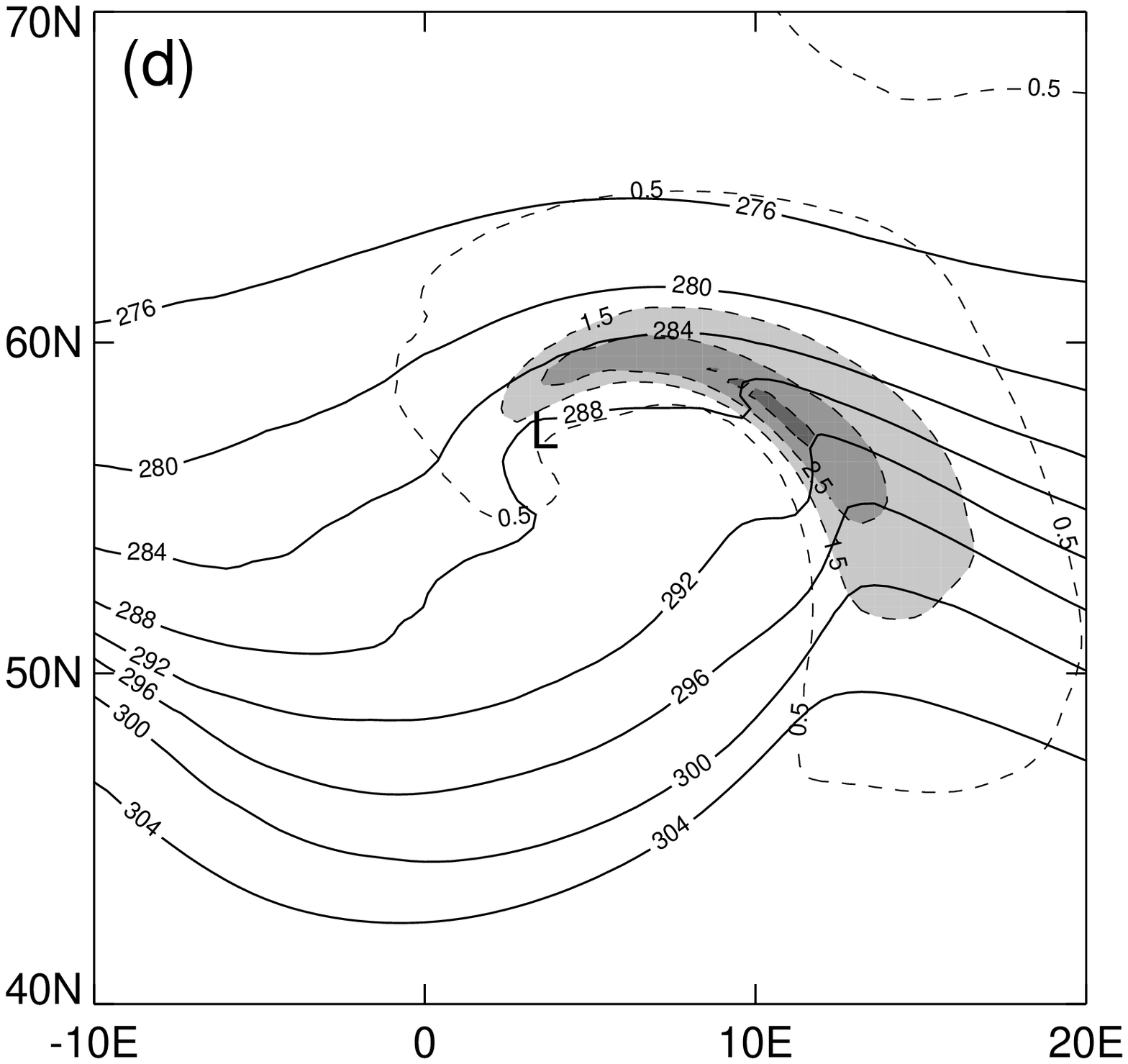}
\caption{Boundary layer depth-averaged potential vorticity (dashed contours, PV Units
$=10^{-6}$~K~kg$^{-1}$~m$^2$~s$^{-1}$) and near-surface potential temperature (solid contours at $4$~K intervals) for (a) the \cite{AdaBHP06} control experiment, (b) the \cite{Bea07} control experiment, (c) the \cite{Bea07} experiment without turbulent heat fluxes and (d) the \cite{Bea07} experiment with meridional SST gradient. Panel (a) is after $6$~days, whilst panels (b)--(d) are after $48$~hours, when the minimum surface pressure is approximately equal in all experiments. Values greater than $1.5$~PVU are shaded and the zero contour is omitted for clarity. `L' denotes the position of the low centre.}\label{fig-pv}
\end{figure*}

The choice of sea-surface temperature (SST) in \cite{Bea07} has considerable effect on the evolution of the boundary layer. The low SST relative to the overlying air in the warm sector gives rise to a shallow (as little as $50$~m in places, but on average about $400$~m) and highly stable boundary layer. The large negative surface heat fluxes in this region lead to the PV budget being dominated by the surface heat-flux term (term~$2$ in Equation~$11$ of \cite{Bea07}); i.e.,
\begin{equation}\label{eq-pv}
\left[\frac{DP}{Dt}\right] \approx -\frac{\xi_h H_{\rm s}}{\rho_0 h^2}
\end{equation}
where $P$ is the Ertel--Rossby potential vorticity, $\xi_h$ is the vertical component of absolute vorticity on the boundary layer top, $H_{\rm s}$ is the surface heat flux, $\rho_0$ is the density, $h$ is the boundary layer depth and the square brackets indicate a depth-average over the boundary layer, after \cite{CooTB92}.
It is this PV which is seen in his Fig.~$4$(b) (our Fig.~\ref{fig-pv}(b)), rather than the PV generated by the baroclinic mechanism which is the focus of \cite{AdaBHP06}.

The importance of the surface heat-flux pattern under conditions of a horizontally uniform SST field is demonstrated by Fig.~\ref{fig-pv}(c), where we have repeated the control experiment of \cite{Bea07} without any turbulent heat fluxes. This makes the simulation closer to that of \cite{AdaBHP06}, since their study focussed on the drag, and so their boundary layer scheme only parameterised momentum transfer and had no turbulent heat fluxes present. Comparing Fig.~\ref{fig-pv}(c) to Fig.~\ref{fig-pv}(b) shows that the PV is now located to the north and east of the cyclone (between $5-15$E, $50-60$N), confirming that
the boundary layer PV in the \cite{Bea07} control run is predominantly generated by turbulent heat
fluxes. The PV generated by heat fluxes remains close to the surface and ahead of the cyclone centre (shown between $10-20$E, $40-50$N in Fig.~\ref{fig-pv}(b)) during the course of the \cite{Bea07} simulation; i.e., it is not vented from the boundary layer. It therefore never reaches a position above the low centre and cannot prevent communication between upper- and lower-level anomalies.

\citeauthor{Bea07}'s \citeyearpar{Bea07} conclusion that the heat-flux generated PV is not dominant in the spin-down process is therefore justified. But the results are not contradictory to those of \cite{AdaBHP06}. Indeed, when the turbulent heat fluxes are switched off, the cyclone shows a slight filling of $2$~hPa after $48$~hours, which is consistent with the results of the PV inversion in \cite{Bea07} that the PV generated by the heat fluxes acts to deepen the cyclone.

\begin{figure}[tbhp]
\centering
\includegraphics[width=\columnwidth]{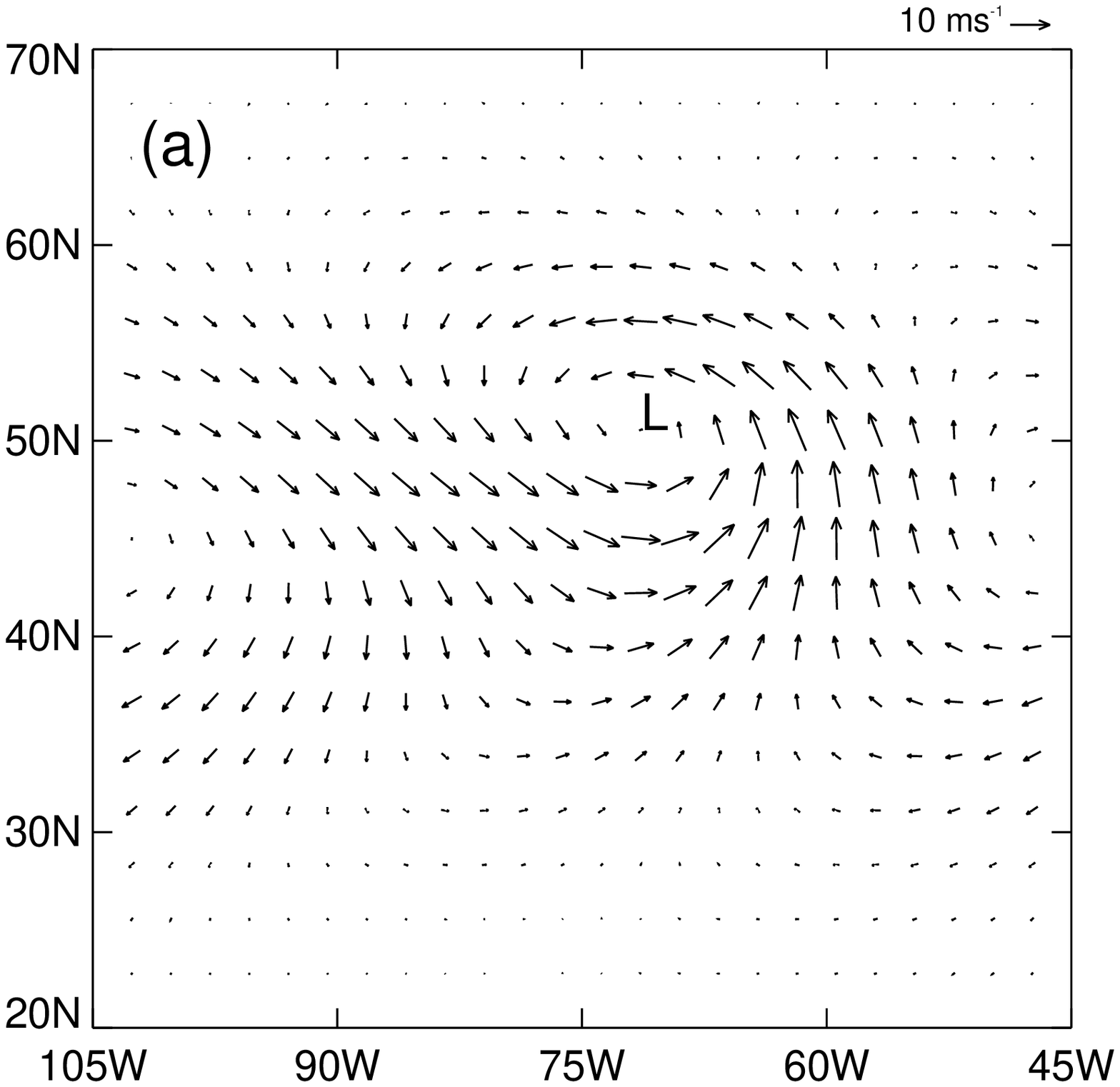}
\includegraphics[width=\columnwidth]{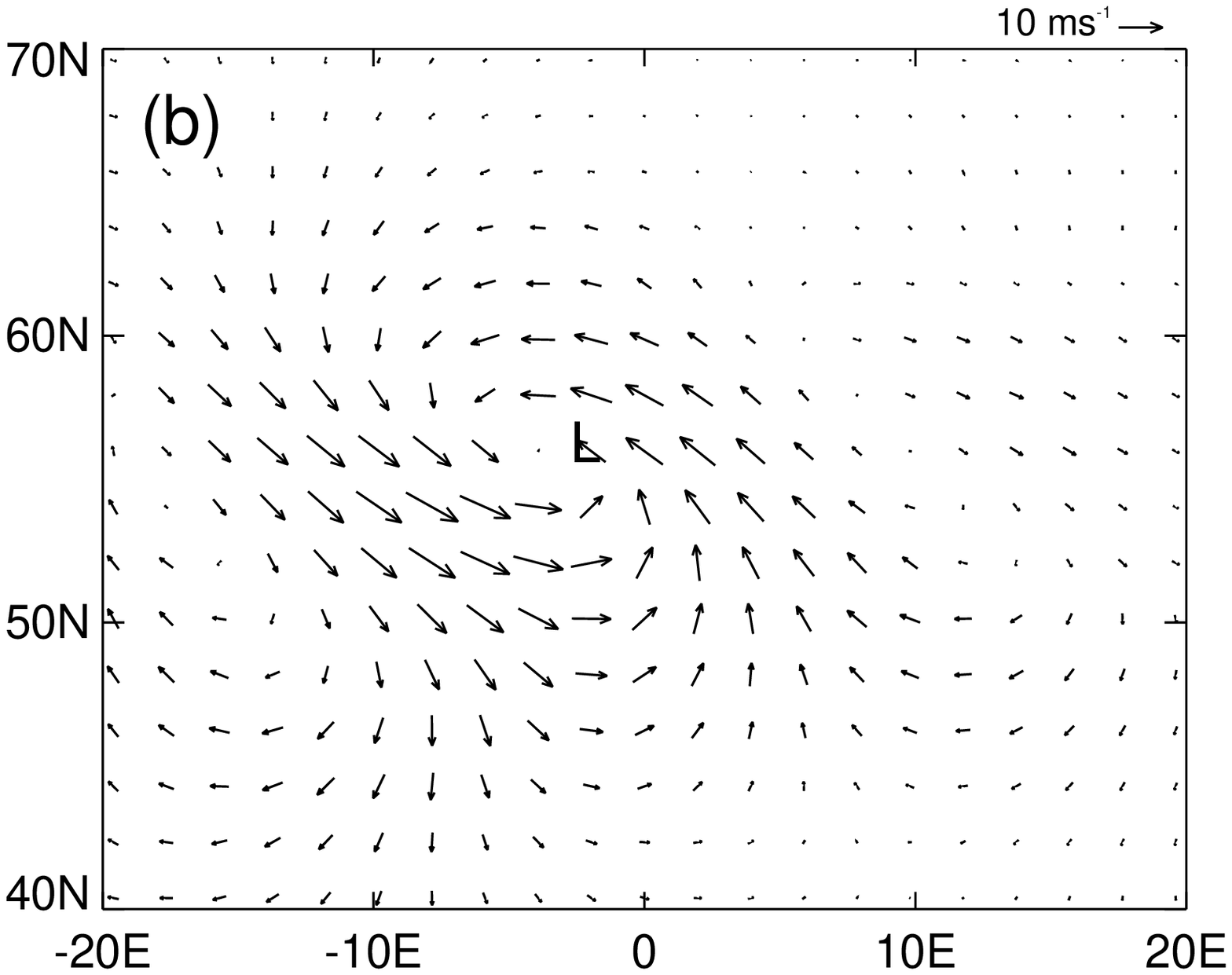}
\caption{$50$~m wind vectors for (a) the \cite{AdaBHP06} control experiment after $4$~days, (b) the \cite{Bea07} control experiment at $24$~hours, when the minimum surface pressure is approximately equal. `L' denotes the position of the low centre. For brevity, we do not show results of the \cite{Bea07} experiment without turbulent heat fluxes or with a meridional SST gradient, as they are very similar to (b).}\label{fig-llj}
\end{figure}

We consider now the low-level jet seen in the simulations of \cite{Bea07}. Formed by a reversal of the north-south temperature gradient generating an easterly wind shear, this cold air wraps around the cyclone centre, producing a cold conveyor belt. This provides ideal conditions for generation of large surface stress. The location of maximum surface stress, or equivalently of the friction velocity, is then found to have a large impact on cyclone development, consistent with the Ekman pumping mechanism. Such a low-level jet is not apparent in the LC1 lifecycle simulations of \cite{AdaBHP06}. Figure~\ref{fig-llj} shows the low-level winds in both simulations at an early stage of development.
If these low-level winds are assumed to lie within the surface layer, the surface stress is given by the bulk aerodynamic formula:
\begin{equation}
\boldsymbol{\tau}_{\rm s}=\rho_0 C_{\rm D} |\bf{v}_1|\bf{v}_1
\end{equation}
where $\boldsymbol{\tau}_{\rm s}$ is the horizontal surface stress vector, $C_{\rm D}$ is the drag coefficient and $\bf{v}_1$ is the horizontal wind vector on the lowest model level. The surface stress exerted on the cyclone is therefore proportional to the square of the low-level wind. It is noticeable that the strongest wind-speeds in Fig.~\ref{fig-llj}(b) are to the southwest of the low centre, between $-15$ and $-5$E, $50-55$N, whereas in Fig.~\ref{fig-llj}(a) they are to the southeast, within the warm sector ($65-60$W, $45-50$N). Therefore, in the \cite{AdaBHP06} experiment, the strongest winds, and therefore surface stress, are in a region of horizontal temperature gradients and hence significant PV generation. However, in \cite{Bea07} the low-level jet wrapping around the cyclone centre enhances wind-speeds to the southwest of the low, making the largest surface stress in a region of small horizontal temperature gradients and hence little PV generation.

By imposing a meridional SST gradient on the \cite{Bea07} experiment, we see both the \cite{AdaBHP06} and \cite{Bea07} mechanisms at work in the same cyclone. The boundary layer depth-averaged PV, shown in Fig.~\ref{fig-pv}(d), is now located to the north and east of the cyclone centre, similarly to Fig.~\ref{fig-pv}(a) and consistent with Figs.~$4$, $5$ and $11$(c) of \cite{AdaBHP06}.

\begin{figure*}[tbhp]
\centering
\includegraphics[width=\columnwidth]{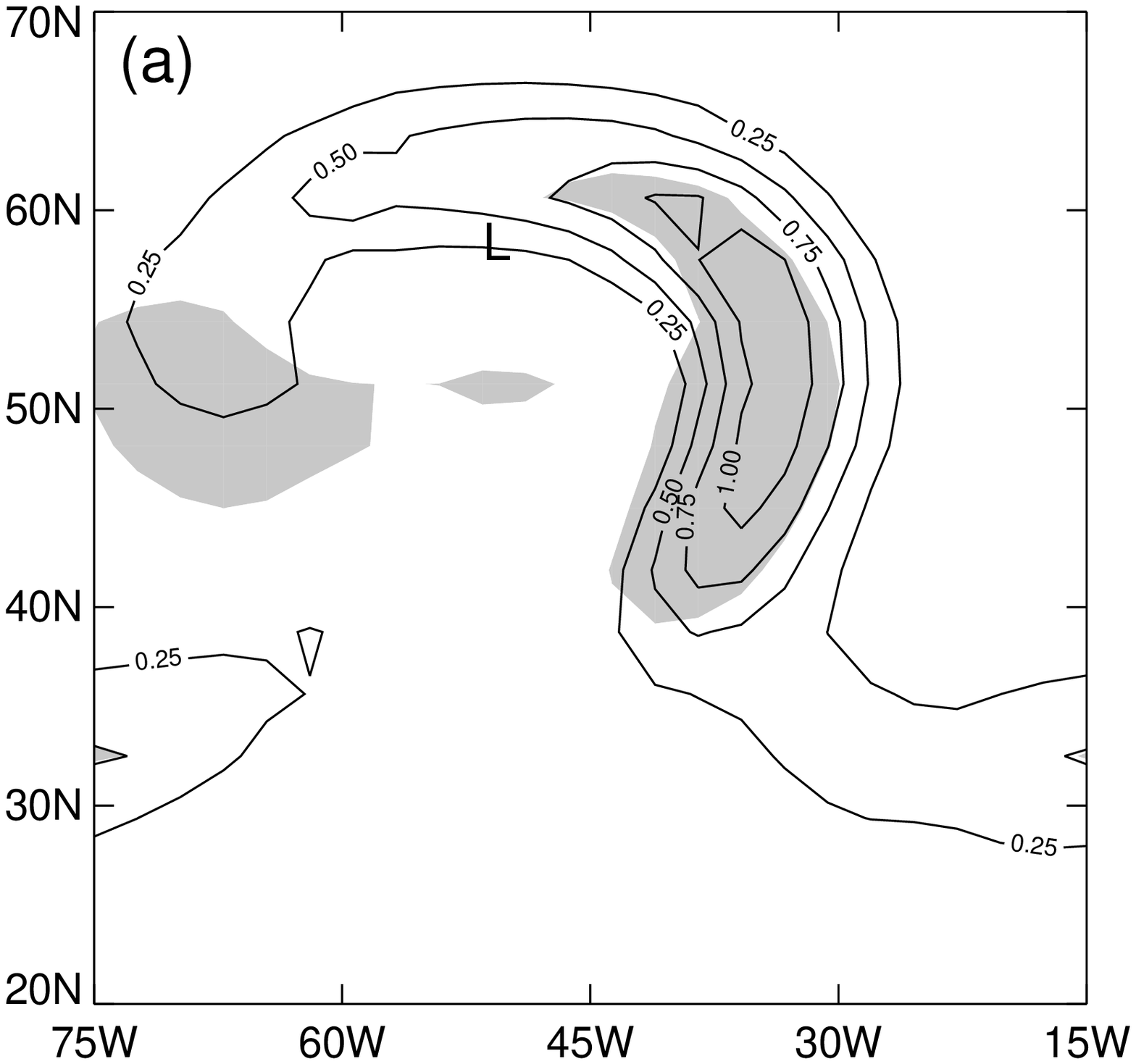}
\includegraphics[width=\columnwidth]{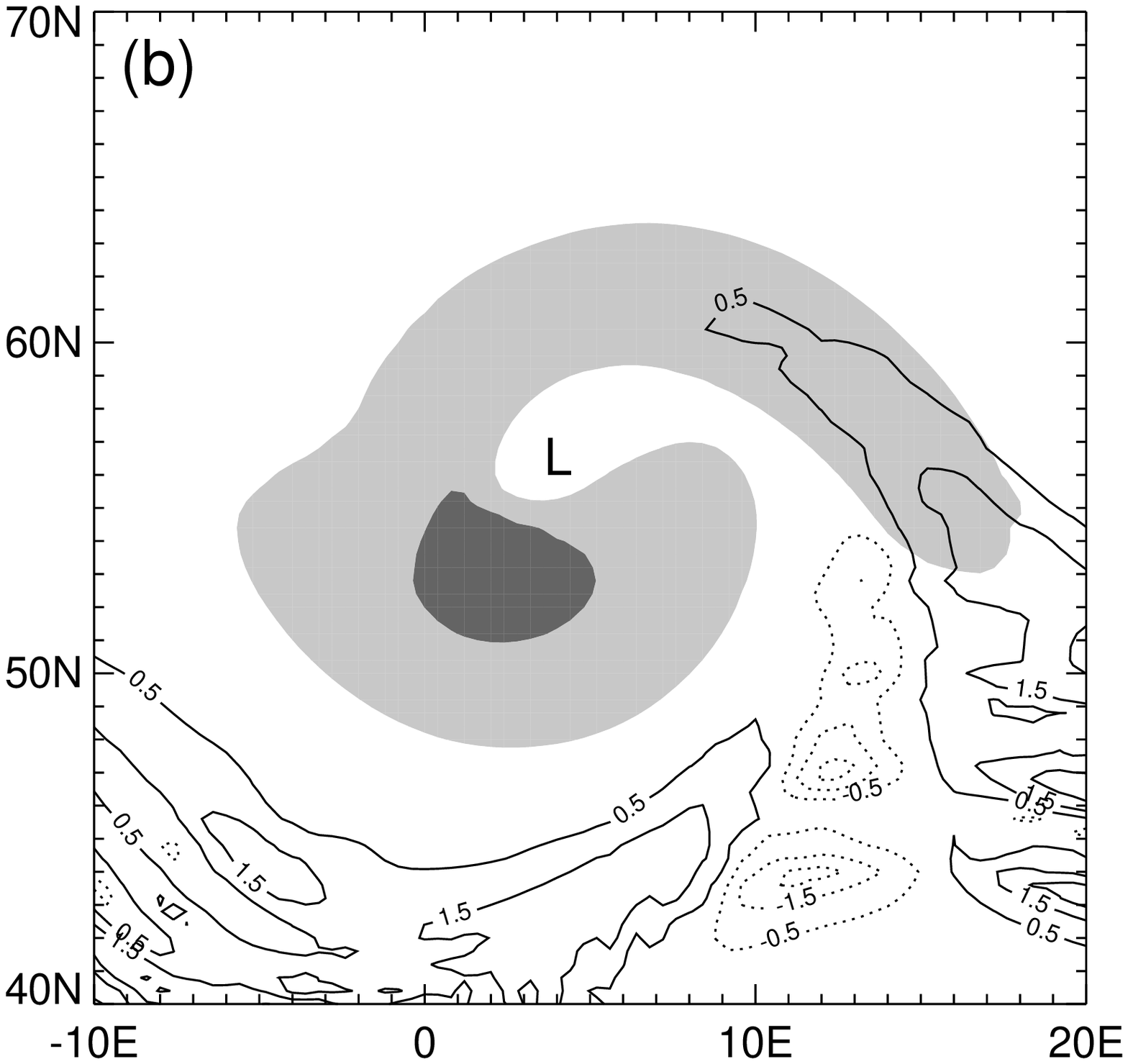}
\includegraphics[width=\columnwidth]{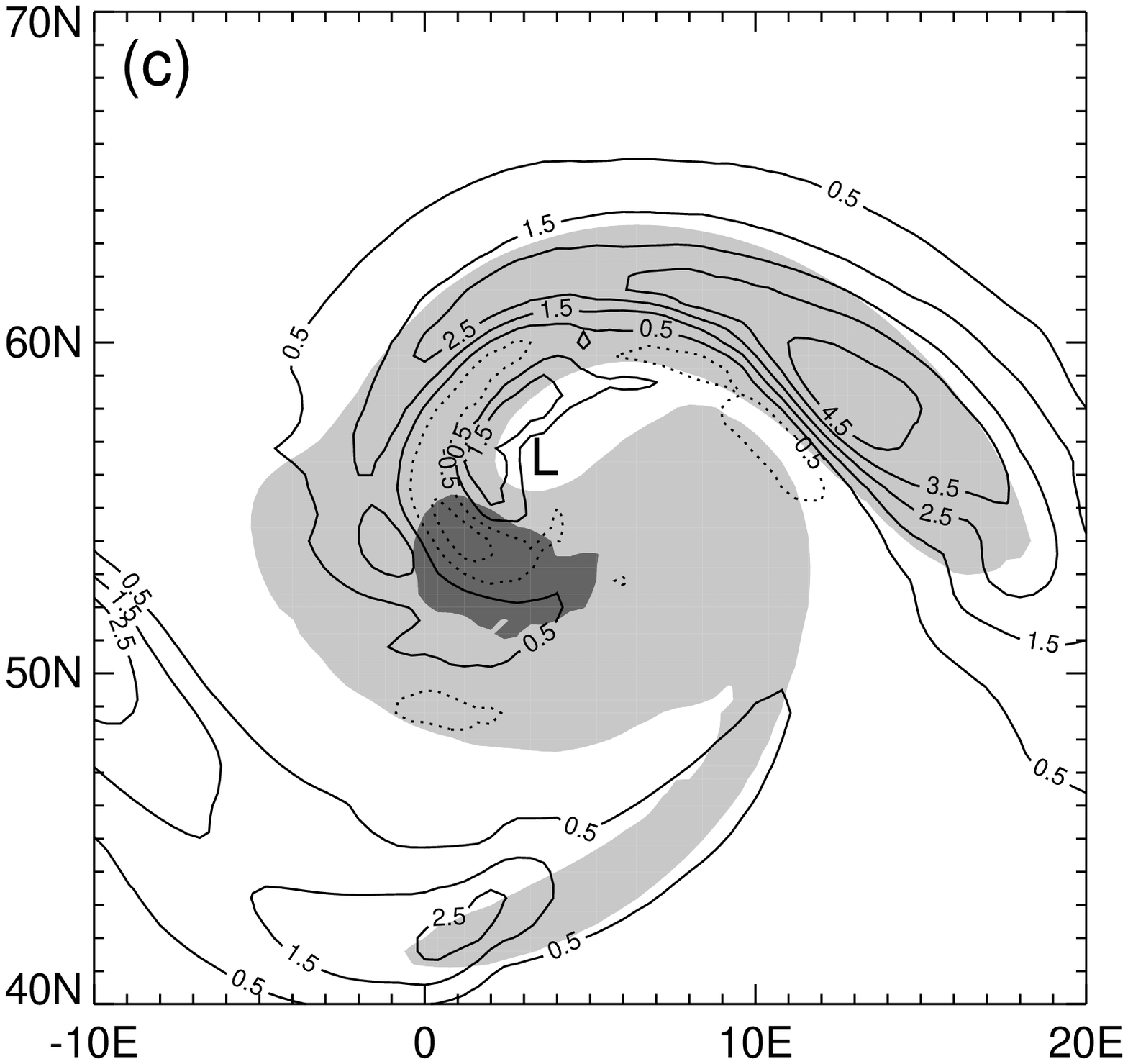}
\includegraphics[width=\columnwidth]{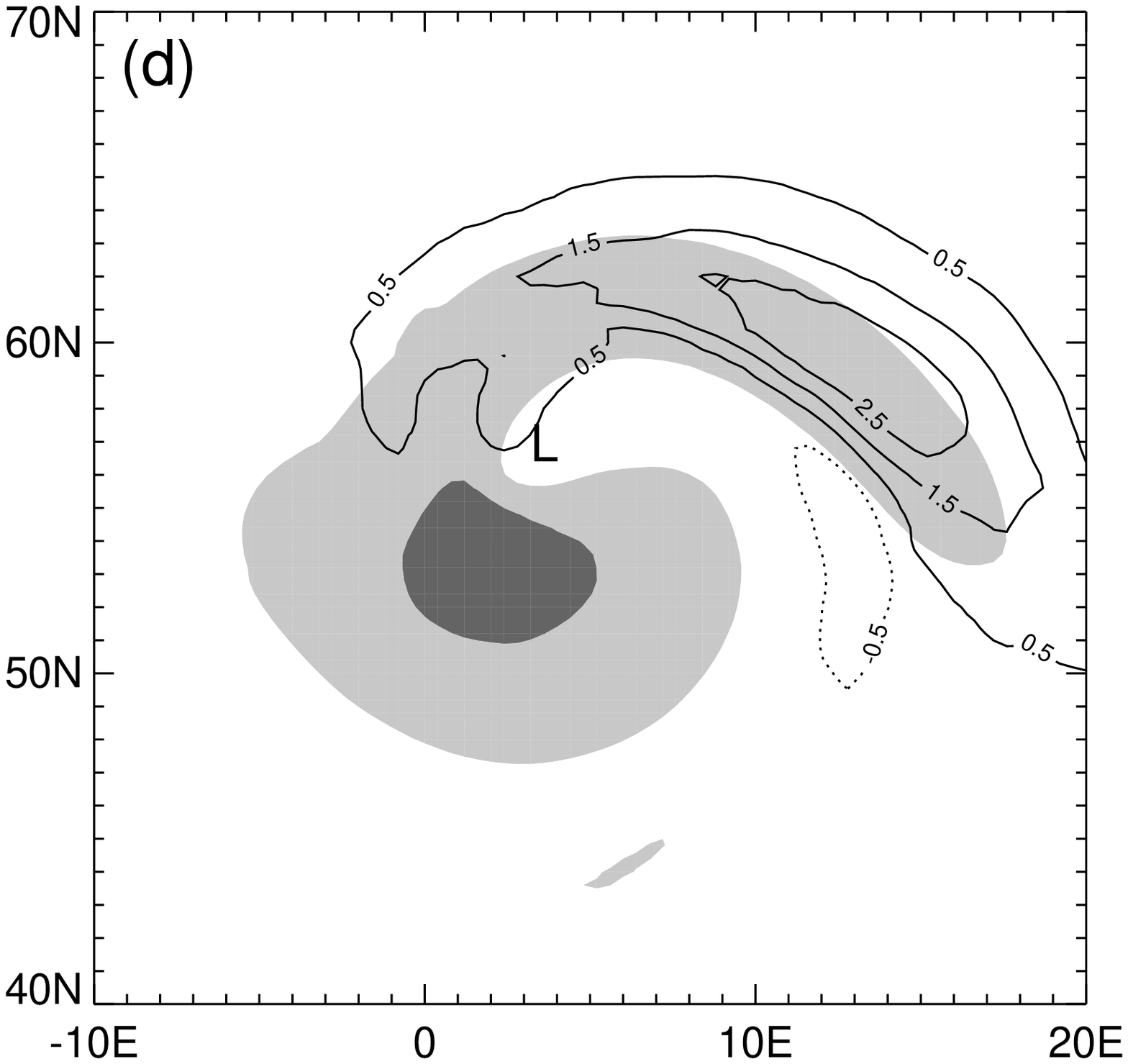}
\caption{Boundary layer depth-averaged potential vorticity
generation by the baroclinic mechanism (PV Units per day, negative values dotted, zero contour omitted for clarity) plotted over friction velocity (values greater than $0.5$~ms$^{-1}$ light grey, values greater than $1$~ms$^{-1}$ dark grey) for (a) the \cite{AdaBHP06} control experiment, (b) the \cite{Bea07} control experiment, (c) the \cite{Bea07} experiment without turbulent heat fluxes and (d) the \cite{Bea07} experiment with meridional SST gradient. Panel (a) is after $6$~days, whilst panels (b)--(d) are after $48$~hours, when the minimum surface pressure is approximately equal in all experiments. `L' denotes the position of the low centre.}\label{fig-gb}
\end{figure*}

Figure~\ref{fig-gb} may be compared with Figs.~$9$ and $10$
of \cite{AdaBHP06} and Fig.~$4$ of \cite{Bea07}. Fig.~\ref{fig-gb}(d) shows that the modified SST experiment has: (i) significant PV generation from the baroclinic mechanism, consistent with \cite{AdaBHP06} (Fig.~\ref{fig-gb}(a)); and, (ii) high values of friction velocity wrapping around the cyclone centre, consistent with \cite{Bea07} (Fig.~\ref{fig-gb}(b)).
The location and magnitude of the friction velocity is consistent with a low-level jet generating maximum surface stress in the well-mixed boundary layer. As discussed by \cite{PlaB07}, PV generation through the baroclinic mechanism has some reinforcement from Ekman and turbulent heat-flux generation terms. The generation shown in Fig.~\ref{fig-gb}(d) occurs between $5$ and $20$E, $55-65$N, in a region well placed to allow ventilation from the boundary layer. \cite{PlaB07} discuss how this ventilation occurs by the cold conveyor belt at early stages of the lifecycle, transitioning to ventilation by the warm conveyor belt at later stages, as the cyclone wraps up. Once advected out of the boundary layer, the PV  appears as a static stability anomaly above the cyclone centre.

\section{Conclusions}

This note aimed to clarify why two recent papers have provided different emphases for the boundary layer's interaction with an extratropical cyclone. We have shown why the experiments of \cite{Bea07} did not find evidence for the baroclinic mechanism, due to the boundary layer PV distribution being dominated by dynamically unimportant PV generation from surface heat fluxes. We have also shown that the much weaker low-level jet in the cold air stream southwest of the low centre in the experiments of \cite{AdaBHP06} limited the Ekman pumping. Through the addition of a meridional SST gradient to the simulations of \cite{Bea07}, we have produced results which show both mechanisms at work. It is beyond the scope of this note to establish the relative importance of each mechanism. That may, for instance, require the development of techniques for PV inversion {\it within} the boundary layer.

In reality, there is of course a spectrum of mid-latitude cyclones, and it is plausible that each mechanism could be dominant in different types of cyclone. Ekman pumping is a direct mechanism, reducing the angular momentum of a pre-existing barotropic circulation, whilst the baroclinic PV mechanism is somewhat indirect, weakening the growth of a baroclinic wave. Certainly more work is thus required to form a complete understanding of the boundary layer processes in baroclinic cyclones.

\acks We would like to thank Andy Brown of the Met Office, UK, for helpful discussions of the work. I.\ Boutle is supported by NERC CASE award NER/S/C/2006/14273.


\begin{thebibliography}{}

\bibitem[Adamson {\em et~al.\ }(2006)]{AdaBHP06}
Adamson, D.~S., Belcher, S.~E., Hoskins, B.~J., and Plant, R.~S. (2006).
\newblock Boundary-layer friction in midlatitude cyclones.
\newblock {\em Q. J. R. Meteorol. Soc.}, {\bf 132}, 101--124.

\bibitem[Beare(2007)]{Bea07}
Beare, R.~J. (2007).
\newblock Boundary layer mechanisms in extratropical cyclones.
\newblock {\em Q. J. R. Meteorol. Soc.}, {\bf 133}, 503--515.

\bibitem[Cooper {\em et~al.\ }(1992)]{CooTB92}
Cooper, I.~M., Thorpe, A.~J., and Bishop, C.~H. (1992).
\newblock The role of diffusive effects on potential vorticity in fronts.
\newblock {\em Q. J. R. Meteorol. Soc.}, {\bf 118}, 629--647.

\bibitem[Plant and Belcher(2007)]{PlaB07}
Plant, R.~S. and Belcher, S.~E. (2007).
\newblock Numerical simulation of baroclinic waves with a parameterized
  boundary layer.
\newblock {\em {J. Atmos. Sci.}} In press.

\bibitem[Shapiro and Keyser(1990)]{ShaK90}
Shapiro, M.~A. and Keyser, D. (1990).
\newblock Fronts, jet streams and the tropopause. Pp 167--191
\newblock in {\em Extratropical Cyclones: The Erik Palm\'en Memorial Volume}. Amer. Meteorol. Soc: Boston.

\bibitem[Thorncroft {\em et~al.\ }(1993)]{ThoHM93}
Thorncroft, C.~D., Hoskins, B.~J., and McIntyre, M.~E. (1993).
\newblock Two paradigms of baroclinic-wave life-cycle behaviour.
\newblock {\em Q. J. R. Meteorol. Soc.}, {\bf 119}, 17--55.

\end{thebibliography}
\end{document}